\documentstyle[aps,twocolumn,epsfig]{revtex}
\begin{document}
\twocolumn[\hsize\textwidth\columnwidth\hsize\csname @twocolumnfalse\endcsname
\title{Nonlocal interactions prevent collapse in Bose-Einstein gases}

\author{V\'{\i}ctor M. P\'erez-Garc\'{\i}a,}

%\begin{instit}
\address{
Departamento de Matem\'aticas, Escuela
 T\'ecnica Superior de Ingenieros Industriales\\
Universidad de Castilla-La Mancha, 13071 Ciudad Real, Spain
}
%\end{instit}

\author{V. V. Konotop}

%\begin{instit}
\address{
Department of Physics and Center of Mathematical Sciences,
University of Madeira,
 Pra\c{c}a do Munic\'{\i}pio, P-9000 Funchal, Portugal
 }
%\end{instit}

\author{Juan J. Garc\'{\i}a-Ripoll}

%\begin{instit}
\address{
Departamento de Matem\'aticas, Escuela
 T\'ecnica Superior de Ingenieros Industriales\\
Universidad de Castilla-La Mancha, 13071 Ciudad Real, Spain
}
%\end{instit}

\date{\today}

\maketitle

%%%%%%%%%%%%%%%%%%%%%%%%%%%%%%%%%%%%%%%%%%%%%%%%%%%%%%%%%%%%%%%%%%%

\begin{abstract}

We study the effect of nonlocality on the collapse properties of a
self-focusing Nonlinear Schr\"odinger system related to Bose-Einstein condensation problems.
 Using a combination of 
moment techniques, time dependent variational methods and numerical simulations 
we present evidences in support of
the hypothesis that nonlocal attractively interacting condensates cannot collapse 
when the dominant interaction term is due to finite range interactions.
Instead there apppear oscillations of the wave packet with a localized component 
whose size is of the order of the range of interactions.
 We discuss the implications of the results to collapse phenomena 
in negative scattering length Bose-Einstein
condensates.

\end{abstract}

]

%%%%%%%%%%%%%%%%%%%%%%%%%%%%%%%%%%%%%%%%%%%%%%%%%%%%%%%%%%%%%%%%%%%%%%

%\narrowtext

\section{Introduction}

The problem of collapses in nonlinear wave equations has been studied
extensively from the mathematical point of view \cite{SIAM}.  In the collapse phenomenon
the amplitude of a physical quantity becomes infinite
at a particular time and usually the mathematical model is no more representative of
the physics of the problem.  This is the case in many particular
instances of collapses in plasma physics \cite{plasmas}, nonlinear optics
\cite{Saturable}, and many other examples.

 In the context of Bose-Einstein condensation there has been recent interest on the analysis of collapse problems. Condensates near $T=0 K$ are modeled by a multi-dimensional Nonlinear Schr\"odinger equation called the Gross-Pitaevskii (GP) equation \cite{Dalfovo}. In this framework, the interaction between the constitutive bosons inside the condensate is defined in terms of the ground state scattering
length $a$. When $a>0$ the interaction between the particles
in the condensate is repulsive which reflects into a positive self-interaction term in the GP equation, whereas for $a<0$, the interaction
is attractive and self-interaction leads to collapse in dimension higher or equal than two 
\cite{Baym,Perez-Garcia97,Japan}.
It is thought that the practical implication is that, in this case, the
 realization of the condensate is limited
by a critical number of particles \cite{Hulet}
above which the condensate is unstable and destroyed
by the collapse phenomenon \cite{Hulet,Baym,Perez-Garcia97,Japan}.
This is the reason why most BEC experiments use gases with positive scattering
length since negative scattering length condensates seem to be limited to 
a very small number of particles. 

In spite of collapse, negative scattering length condensates have some
peculiarities which make them interesting.  One of them is that if the
trap is removed in one direction, the attractive interaction yields (for
a particular number of particles) to a self-confined stationary state as it
has been shown in a previous work \cite{Perez-Garcia98}.  In that case,
the cloud as a whole behaves as a particle and can be controlled by
acting on it with external fields.  These solutions can be properly
called solitons since they appear as a consequence of self-interaction
and not merely because of the effect of an external potential. These solitonic
condensates would thus be more controllable than usual repulsive condensates. 

 There is another reason why one would be interested in those condensates. Although
it is thought that these condensates are unstable, this fact has not been really tested experimentally.
What Rice group experiments established \cite{Hulet} was that it is not possible to {\em generate}
a large enough negative scattering length condensate. This problem was analyzed in more depth in Ref.
\cite{Hulet2} were the authors used the two fluids model
 to analyze the condensation process. They found a natural bound on the number of particles attainable by the evaporative cooling method which is used to grow the condensate.

 However, condensate growth is not the only way to generate large negative scattering length
condensates. Recent experimental results\cite{Inouye98} show that
it is possible to use Feshbach resonances to continuously
detune the value of $a$ from positive to negative values,
by means of external magnetic fields. This provides new interest to
the analysis of attractively interacting condensates since it would provide an 
alternative way to generate them. Another way these large condensates could be obtained is
by overlapping two
subcritical negative scattering length  stable Rb or Cs condensates as
proposed in Ref. \cite{Michinel}. In that paper it has been shown that fringe systems generated by interference can be used to stabilize large overlapping negative scattering length
condensates. However, no matter how one generates the supercritical  negative scattering length condensate the question 
is: what would be its subsequent dynamics?

To make a complete analysis of that point, i.e. dynamics of an
already generated supercritical condensate near $T=0 K$; one should take into 
account at least the effect of new
terms correcting the usual Gross-Pitaevskii equation.
 Although there is the possibility of considering the problem in more 
general frameworks such as the two fluids model or even the full quantum problem we will concentrate here on simpler 
formalisms which are already very rich as will be shown later. Between the possible
corrections to GP equation are those due to four-particle collisions, which
 were considered in Refs. \cite{Schlap,Michi2}. However the real relevance of those
terms is doubtful since there is no idea on what could be
the value of the coefficient of the higher order correction to the cubic nonlinear term.
Moreover it can be seen that they must be very small \cite{Michi2} and
probably losses terms are quite relevant before this term becomes
appreciable.  Finally it is not clear that one could perform a Taylor
expansion of the nonlinear interaction functional around the minimum
point in that situation, a requisite which must be satisfied if the true interaction
potential is to be approximated by a polynomial on $|\psi|$.

 On the other hand, there is a characteristic of the condensate self-interaction which is neglected 
in the usual treatment: the fact that the true interaction potentials are nonlocal. This is the way the self-interaction appears
 in the mean field formulation of the Hartree-Fock problem
for the Boson system \cite{LP,Dalfovo}. In this paper we study the effect of nonlocal interaction terms
on the collapsing properties of the Gross-Pitaevskii equation. 
 We do not claim that this will be the most relavant physical contribution in all the possible situations
but we believe that it could be in some of them due to the many different atomic species which are known to possess negative
scattering length. It is thus interesting to understand what is the isolated 
effect of nonlocal nonlinear terms. Our interest will be twofold:
firstly from the point of view of nonlinear science we are interested in understanding what
is the effect of such term on the collapse problem; secondly, we will consider
the problem from the viewpoint of Bose-Einstein condensation phenomena.

The study of nonlocal nonlinear wave equations is nontrivial and
only recently there have been several studies considering particular
cases in the contexts of nonlinear optics
\cite{Alfimov,Vanin,Abe,Akhmendiev}, Quantum Mechanics \cite{Filipov}
and other nonlocal wave equations \cite{Vazquez2}.  Presently only 
few rigorous results and some
qualitative estimates are available for very particular cases. To our
knowledge, the only related field where these nonlocal Nonlinear
equations (in particular of Nonlinear Schr\"odinger type) have been
studied in detail and with mathematical rigor is scattering theory, but
the results are completely disjoint with our purpose here
\cite{Ginibre}.  Finally, in the context of Bose-Einstein condensation
there has been only a work known to us, in which non locality effects are
considered \cite{Parola}. The authors use the gradient expansion to obtain a local
interaction energy and then perform a classical
time independent variational analysis with Gaussian ansatz to estimate
the possibility that nonlocality enhances stability conditions.
An important conclusion made by the authors is that 
nonlocal interactions could prevent collapse. However, 
the results obtained so far
are based on qualitative arguments (the gradient expansion is not 
applicable in the region where the narrowing of the wavepacket is stopped) and leave open the question about the dynamics
of the nonlocal interacting condensate. In our paper
we complement static results with mathematically rigorous analysis of the 
wavepacket width evolution, time dependent variational analysis,
 numerical simulations and the
analysis of the strongly nonlocal limit. Using analysis of the 
dynamics on time we can say on more firm ground that nonlocal 
interactions of a rather general (but probably not any) type 
prevent collapse.

Our detailed plan is as follows.  In Sec.  \ref{II} we present the model
equations with the nonlocal term.  A formal analysis of the
strongly nonlocal  limit analysis 
and wave packet width evolution using the moment method is
done in Sec. \ref{infty}. In Sec.  \ref{III} we present some
analytical approximations to obtain a qualitative description of the
collapse dynamics in all the regimes. In Sec. \ref{numerics} 
we present the results 
of numerical simulations of the full model and compare the 
results with analytical predictions of previous sections. Finally, 
in Sec.  \ref{IV} we summarize our conclusions.

\section{Mathematical Model}
\label{II}

The usual theoretical model used to describe a system of 2-body
interacting bosons of mass $m$, with a fixed mean number $N$, trapped in
a parabolic potential $V({\bf r})$ is the
Gross-Pitaevskii (GP) equation \cite{Dalfovo}
\begin{eqnarray}
\label{GPE}
 i \hbar \frac{\partial \Psi}{\partial t} & = & - \frac{\hbar^2}{2 m}
\nabla^{2}\Psi + V({\bf r})\Psi \nonumber \\
  & + & U \left[\int K({\bf r}-{\bf r'}) |\Psi({\bf r'})|^2 d{\bf r'} \right]\Psi,
\end{eqnarray}
where $U = 4 \pi \hbar^2 a/m$
characterizes the 2-body interaction and has been extracted from the
kernel $K$ for later convenience. The normalization
for $\Psi$ is $N = \int |\Psi|^2 \ d^3 \bf{r},$
and the trapping parabolic potential is given by
$V({\bf r}) = \frac{1}{2} m \nu^2 \left( \lambda_x^2 x^2 +
\lambda_y^2 y^2 + \lambda_z^2 z^2 \right),$
as discussed e.g. in Ref. \cite{Perez-Garcia97}. Here we have kept the
nonlocal interaction as it appears in the Hartree-Fock theory.
Usually the kernel function is taken to be a delta function because of the low energy of the interaction process.  One can easily recover that case by imposing that
$K({\bf{r}}-{\bf{r}'}) =
\delta({\bf{r}}-{\bf{r}'})$. In our treatment we will keep the nonlocal nature of the kernel and 
show that it gives rise to many new effects.

Let us write the equations
 in a natural set of units which for our problem is built up from two
scales: the trap size (measured by the width of the linear ground state), $a_{0}=\sqrt{\hbar/m\nu}$,
and its period, $1/\nu$.  The new variables are $(x_0,y_0,z_0) = (x,y,z)/a_0, \tau = \nu t, \psi({\bf r}_0) = \Psi({\bf r}) \sqrt{a_0^3/N}$. With these definitions the equation
simplifies to
\begin{eqnarray}
\label{GPEredu}
 i \frac{\partial \psi}{\partial \tau} & = & - \frac{1}{2}
\nabla^{2}_0\psi +
\frac{1}{2} \left(\lambda_x^2 x_0^2 + \lambda_y^2 y_0^2 + \lambda_z^2 z_0^2 \right) \psi \nonumber \\
  & + & U \left[\int K({\bf r_0}-{\bf r'}) |\psi({\bf r'})|^2 d{\bf r'}
\right]\psi,
\end{eqnarray}
being $U=4\pi Na/a_{0}$.
The particular shape of the kernel function
could depend strongly on the energy of the interaction and the geometry of the
molecules involved and probably it is very difficult to know its precise shape except
for the case of atomic hydrogen, in which BEC has been recently achieved
\cite{Hydrogen}.

 In this paper we are interested in basic general qualitative results and we do not try to model 
the specific details of the interaction for any particular atom. This is why we will concentrate in the simplest case where the kernel depends on one parameter
$\epsilon$, related to the kernel ``size" which charaacterizes the range of interactions, such that
\begin{equation}
\lim_{\epsilon \rightarrow 0} K_{\epsilon}({\bf{r}}) = \delta({\bf{r}})
\end{equation}
Moreover, we will analyze molecules (and thus the interaction) which are spherically symmetric. The main implication of this fact is that the kernel depends  only on
the distance between two atoms, $K(|{\bf r}-{\bf r'}|)$.
The question we will address in what follows is the possibility of blow-up
with regular kernels $K_{\epsilon}$. This is a very intricate
mathematical problem, which even in the simplest $\delta$-case  (cubic
nonlinearity) is not completely solved, and only some
 estimates on collapse conditions exist. Thus, we will not try to provide complete
rigorous proofs but only to join numerical simulations and  approximate
analytical techniques to understand the problem.

\section{Strongly nonlocal limit}
\label{infty}

Let us start with some physical arguments showing that the nonlocality
can prevent collapse. Although it is not essential the algebra is simplified by assuming 
 cylindrically $(\lambda_x = \lambda_y = 1)$ or spherically symmetric $(\lambda_x = \lambda_y = \lambda_z = 1)$ traps. We notice that there exist at least two
integrals of motion of (\ref{GPEredu}):
\begin{mathletters}
\begin{eqnarray}
\label{N}
N& =& \int |\psi|^2\,d{\bf r} ,\\
\label{H}
H & = & \int \left(|\nabla \psi|^2+J({\bf
r})|\psi|^2+\rho^2|\psi|^2\right)\,d{\bf r},
\end{eqnarray}
\end{mathletters}
where $ \rho = |{\bf r}|$ and 
\begin{equation}
J({\bf r})=U\int K({\bf r}-{\bf r}')|\psi({\bf r}')|^2\,d{\bf r}'.
\end{equation}
$N$ is the number of particles because of the normalization imposed on $\psi$ and
$H$ is associated with the energy of the condensate.

Let us now consider the situation ``near" the
collapse assuming that the wave function is strongly localized. More
precisely, it will be assumed that the localization region of the wave
function $\ell$ is much less than the range of the nonlocal
interactions $\epsilon$, $\ell \ll\epsilon$\cite{com}. Also it wwill 
be assumed that the kernel funnction is nonsingular at ${\bf r}\to 0$. 
Then in the leading approximation one can approximate
\[
J({\bf r})=UNK({\bf r})+O(\ell /\epsilon)
\]
Making the natural supposition that if the collapse occurs it happens at ${\bf
r}=0$ (i.e. at the minimum of the confining potential) one finds the following 
linear equation for the wave function
\begin{equation}
\label{linear}
i\frac{\partial\psi}{\partial
\tau}=-\frac{1}{2}\nabla^2_0 \psi+UNK(0)\psi
\end{equation}
In Eq. (\ref{linear}) we have neglected the confining potential, since it is of order of
$\ell^2\nu \ll 1$.

As it is evident, Eq. (\ref{linear}) does not display collapse. Moreover the
dispersion will lead to spreading out of any initially localized wave
packet.

On the other hand if the condensate width is much bigger than the range of
interactions $\ell \gg \epsilon$, the $\delta$-function limit is applicable and the dominant nonlinearity will lead to collapse. 
Thus we conclude that there must exist two different tendences: spreading for $\ell \ll \epsilon$ and narrowing for $\ell \gg \epsilon$ of the wave packet. This must lead to oscillations of the wave packet width between different scales, the smaller one being of
the order of the interaction range. 
As a matter of fact the  limit $\ell \ll \epsilon$ is just opposite to the limit of
local interactions and can be interpreted as the case of infinite range
interactions, $K({\bf r})\equiv K(0)$. So we see that there is a scale at which the interaction
is attractive and a local scale at which it is repulsive. It seems natural that this will 
be reflected on oscillation in the wavefunction dynamics, a fact which we will analyze later.

 It is also possible to argue that the
collapse does not occur in this system using the more conventional language of momenta
\cite{Japan,Zakharov}. To this end  let us define the mean squared width of the wave packet
\begin{equation}
\label{momenta}
\langle\rho^2\rangle=2\pi
(n-1)\int_0^{\infty}|\psi(\rho)|^2\rho^{n+1}d\rho
\end{equation}
where $n=2,3$ is the spatial dimension. Then it is a straightforward algebra to obtain
\begin{eqnarray}
\label{EvolRho}
\frac{d^2}{d\tau^2}\langle\rho^2\rangle &= 
& 2H-4\lambda^2\langle\rho^2\rangle-4\pi
(n-1) U \nonumber \\
& \times & \int_0^{\infty}|\psi|^2\left[J(\rho)+\frac{1}{n-1}\rho J'(\rho)\right]
\rho^{n-1}d\rho
\end{eqnarray}
In the limit of infinite range of interactions (or very localized wavefunctions) one finds
\begin{equation}
\label{EvolInf}
\frac{d^2}{d\tau^2}\langle\rho^2\rangle=
2H-4\lambda^2\langle\rho^2\rangle+\frac{1}{\pi (n-1)} UN^2
\end{equation}
The solution of Eq. (\ref{EvolInf}) reads
\begin{equation}
\label{solution}
\langle\rho^2\rangle=\rho_0^2\sin(4 \tau+\phi_0)+ C
\end{equation}
where $C = 2H+UN^2\frac{1}{\pi(n-1)}$, $\rho_0$ and $\phi_0$ are real constants. It follows from the energy conservation law that
\begin{equation}
\label{C1}
C \geq 2\langle \rho^2 \rangle
\end{equation}
 Hence the minimal
value of $\langle \rho^2\rangle$, given by
$C -\rho_0^2$ is always positive $\langle \rho^2\rangle \geq 0$
since it follows from (\ref{C1}) that
$C/4 \geq\rho_0$. This result rules out the possibility that it could exist a collapse in which all 
the wavefunction became concentrated on one particular point (e.g. a delta-like singularity).

\section{Analytical results for the general case}
\label{III}

\subsection{Exact results for the center of mass}

 It has been pointed out in previous works \cite{PRL,Juanjopub}  that
the center of mass defined by
\begin{equation}
\label{CentMass}
\langle {\bf r}_0\rangle=\int{\bf r} |\psi|^2 d {\bf r}
\end{equation}
performs harmonic oscillations no matter
what the nonlinear interaction is. This result is also valid for
the nonlocal interaction case under very general conditions as  follows
from Erhenfest theorem extended to this case so that
\begin{mathletters}
\begin{eqnarray}
  \frac{d^2}{d\tau^2}<x_0> + \lambda_x^2 <x_0> =0,
  \label{edo-cm-x} \\
  \frac{d^2}{d\tau^2}<y_0> + \lambda_y^2 <y_0>=0,
  \label{edo-cm-y} \\
  \frac{d^2}{d\tau^2}<z_0> +  \lambda_z^2 <z_0>=0.
  \label{edo-cm-z}
\end{eqnarray}
\end{mathletters}
 To obtain more exact information on this problem a possibility is to use the moment method \cite{nuestro,Porras} in radial symmetry.  Moreover it can be
seen that it does not provide exact results for our case. Even the use of
the uniform divergence approximation \cite{nuestro} is not possible and the only possibility is to restrict to the classical time dependent variational techniques, which we consider in the following
subsection.

\subsection{Time dependent variational formalism}

 Following the standard procedure we first identify a Lagrangian density for
 problem (\ref{GPE}), which is
\begin{eqnarray}
  \label{density} {\cal L} & = & \frac{i\hbar}{2} \left(
  \Psi\frac{\partial \Psi^{\ast}}{\partial t} - \Psi^{\ast}
  \frac{\partial \Psi}{\partial t} \right) + \frac{\hbar^2}{2m} |\nabla \Psi|^2 \nonumber \\ & + &
   V(r) |\Psi|^2 + \frac{U}{2} \int K({\bf r}-
{\bf r'}) |\Psi({\bf r})|^2|\Psi({\bf r'})|^2 d{\bf r'},
\end{eqnarray}
where the asterisk denotes complex conjugation. That is, instead of
working with the Gross-Pitaevskii equation we can treat the action,
\begin{equation}
\label{action}
  S = \int {\cal L} d^3rdt = \int_{t_i}^{t_f} L(t) dt,
\end{equation}
where $t_i$ and $t_f$ are initial and final moments of time, 
and study its invariance properties and extrema, which are in turn
solutions of Eq. (\ref{GPE}).

To simplify the problem, we restrict the shape of the function $\Psi$ to
a convenient family of trial functions and study the time evolution of
the parameters that define that family. A natural choice, which
corresponds to the exact solution in the linear limit ($U = 0$) and
provided quite good results in our previous works
\cite{Perez-Garcia97,PRL} is a $n$-dimensional Gaussian-like function
\begin{equation}
  \label{ansatz}
  \Psi(x,y,z,t)  =  A \prod_{\eta}
  \exp \left\{ \frac{-[\eta-\eta_{CM}]^2}{2w_\eta^2}
    + i \eta \alpha_\eta+ i \eta^2\beta_\eta \right\}.
\end{equation}
where  $A$ (amplitude), $w_\eta$ (width), $\alpha_\eta$
(slope -- speed), $\beta_{\eta}$ (square root of the curvature radius) and
$\eta_0$ (center of the cloud) are free parameters. 
We are considering $n$ as a free parameter that can be set to two or three depending on the dimensionality of the
problem considered. BEC systems are in general three dimensional; however, in certain situations a two dimensional
condensate can be considered a good theoretical model \cite{twodimen} and is easier to compare with numerical simulations of Eq. (\ref{GPE}).

 The procedure for deriving equations for the parameters has been described in previous works \cite{Perez-Garcia97,PRL} and will not be repeated here.

 To go on with the analysis a particular shape must be chosen for the kernel function. We will consider a simple
$n$-dimensional Gaussian kernel of the form
\begin{equation}
K({\bf r}) = \left(\frac{1}{2\pi \epsilon^2}\right)^{n/2} e^{-{\bf r}^2/2\epsilon^2}
\end{equation}
The main equations obtained when computing the evolution equations for these
kind of systems are those related to the width. To do so it is useful to introduce a set of rescaled variables for
time, $\tau = \nu t$, and the widths, $w_\eta = a_0 v_\eta,
(\eta=x,y,z)$. For a $n$-dimensional condensate the equations are found to be
\begin{equation}
\label{widths2bis}
  \frac{d^2v_k}{d\tau^2} + \lambda_k^2v_k  =
    \frac{1}{v_k^3} + \frac{Pv_k}{v_k^2+\delta^2} \prod_{\eta} \frac{1}{\left(v_{\eta}^2 + \delta^2\right)^{1/2}}, \label{vx}
\end{equation}
where $P = \sqrt{2/\pi}Na/a_0$ (strength of the atom-atom interaction) and $\delta = \epsilon/a_0$.
The remaining parameters $\alpha_{\eta},\beta_{\eta}$ obey separate equations but essentially can be computed independetly once the $v_k(t)$ are known. 

The system (\ref{widths2bis}) is generated by the Hamiltonian
\begin{eqnarray}
H & =& \frac{1}{2} \sum_{\eta} \dot{v}_{\eta}^2 + \frac{1}{2} \sum_\eta
\left(\lambda_{\eta}^2v_{\eta}^2+\frac{1}{v_{\eta}^2}\right) \nonumber \\
\label{hamilt} & + & P\prod_\eta \frac{1}{\left(v_{\eta}^2+\delta^2\right)^{1/2}}.
\end{eqnarray}
In terms of the system (\ref{widths2bis}) collapse corresponds to the behavior when 
$v_k\rightarrow 0$. As a matter of fact the explicit expression of the Hamiltonian 
already shown that collapse is prevented by nonlocal interactions (i.e. by nonzero $\delta$)
in the framework of this simple model. Indeed one can easily see that $H$ has a lower bound
$H \geq - P/\delta^3$.

\subsection{2D case: Equilibrium points and minimum width}

\begin{figure}
\epsfig{file=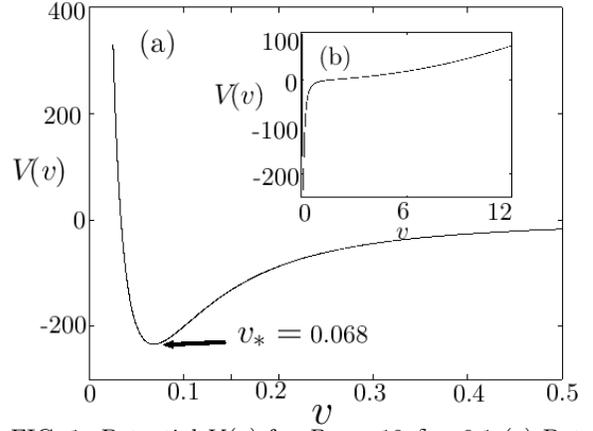,width=7.5cm}
\caption{Potential $V(v)$ for $P=-10, \delta = 0.1$ (a) Detail of the
small scale (b) Large scale}
\label{fig1}
\end{figure}

\begin{figure}
\epsfig{file=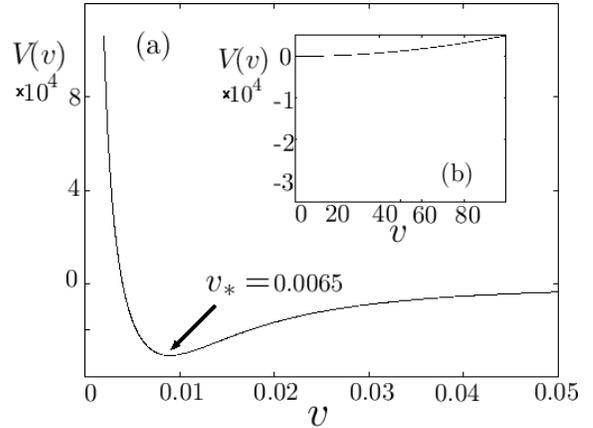,width=7.5cm}
\caption{Potential $V(v)$ for $P=-10, \delta = 0.01$ (a) Detail of the
small scale (b) Large scale}
\label{fig2}
\end{figure}

In the present section we will concentrate on two dimensional
condensates for the sake of comparison with
numerical simulations of Eq. (\ref{GPEredu}), which will be presented in Sec. \ref{numerics}. In this case the equations are
\begin{mathletters}
\label{widths2bisb}
\begin{eqnarray}
  \frac{d^2v_x}{d\tau^2} + \lambda_x^2v_x  =
    \frac{1}{v_x^3} + \frac{Pv_x}{\left(v_x^2+\delta^2\right)^{3/2} \left(v_{y}^2 + \delta^2\right)^{1/2}}, \label{vxbis} \\  \frac{d^2v_y}{d\tau^2} + \lambda_y^2v_y  =
    \frac{1}{v_y^3} + \frac{Pv_y}{\left(v_y^2+\delta^2\right)^{3/2} \left(v_{x}^2 + \delta^2\right)^{1/2}}, \label{vybis}
\end{eqnarray}
\end{mathletters}
Let us consider the case when $v_x = v_y = v$, which corresponds to a cylindrical symmetry 
around the center of the wavefunction. When the solution has the symmetry of the external potential (i.e. $\eta_{CM} = 0$) this corresponds to the usual cylindrically symmetric case since the wavefunction amplitude depends only on ${\bf r}$. In this case the equations are simpler
\begin{equation}
  \frac{d^2v}{d\tau^2} +  v  =
    \frac{1}{v^3} + \frac{Pv}{\left(v^2 + \delta^2\right)^{2}}
\end{equation}
This equation can be obtained from the potential
\begin{equation}
V(v) =  \frac{1}{2} v^2 + \frac{1}{2v^2} + \frac{P}{2 \left(v^2+\delta^2\right)}
\end{equation}
which is plotted in Figs. \ref{fig1} and \ref{fig2} for some particular parameter values.
It can be seen that even when $P$ is negative, corresponding to negative scattering length, the potential is repulsive at the origin so that no blowup is possible.

The case of interest for us corresponds to small $\delta$ values. In that limit the potential has two scales. For $v = {\cal O}(\delta)$
the parabolic term can be neglected since the last two terms are dominant. For $ v = \cal{O}(1) - {\cal O}(1/\delta)$ the last term
can be neglected (at least when $P$ is not large) and the potential is dominated by the parabolic term. 
The existence of two length scales is also clear in 
Figs. \ref{fig1} and \ref{fig2}.

 The equilibrium points $v_*$ are the solutions of the algebraic equations
\begin{equation}
\label{equil}
 v = \frac{1}{v^3} + \frac{Pv}{\left(v^2 + \delta^2\right)^2}
\end{equation}
The $\delta = 0$ case was analyzed in Ref. \cite{Perez-Garcia97}; now the situation
is quite different since collapse is not possible.
In a generic case one can show that there exists only one positive root
of Eq. (\ref{equil}). To this end we introduce a new variable  $z$,
$v/\delta =e^{z}$ and rewrite Eq. (\ref{equil}) in the form
\begin{equation}
\label{equil1}
\sinh(2(z+z_0))\cosh^2 z-\frac{1}{4}Pe^{2z_0} =0
\end{equation}
where $\delta=e^{z_0}$ . The left hand side of this equation is a
monotonic function of $z$ and thus there exists only one real root of
Eq. (\ref{equil1}). Moreover the root is finite for $\delta\neq 0$ what
means that the collapse ($v=0$) is not possible. This result
is in agreement with the consideration of Sec. \ref{infty}.

Let us concentrate on the small $\delta$ case, to do so let us first 
define a new variable $q = (v/\delta)^2 > 0$, so that Eq. (\ref{equil}) becomes
\begin{equation}
\delta^4 = \frac{1}{q^2} + \frac{P}{\left(q + 1\right)^2}.
\end{equation}
Since we are going to deal with small $\delta$ values we can neglect the left hand side \cite{phys}
and obtain after some algebra the only equilibrium point as $q_* = \frac{1}{\sqrt{|P|}-1}$, provided $|P|>1$,
which implies that
\begin{equation}
\label{cero}
v_* \simeq \frac{\delta}{\sqrt{\sqrt{|P|}-1}},
\end{equation}
and is consistent with our previous assumptions on the linear term. The fact that the equilibrium point depends linearly on $\delta$ is
interesting and provides a first estimate of the order of
 magnitude of the turning point of the potential which is then roughly
${\cal O}(\delta)$ for a wide range of initial energies.
We insist that this is a lower bound for the minimum width since it could happen (as it does) that there is only part of the solution 
of Eq. (\ref{GPE}) which tends to collapse and then the contribution of the noncollapsing part will make the width finite. In that case our estimate would roughly correspond to the width of the collapsing peak. 

 If $|P| \approx 1$ Eq. (\ref{cero}) cannot be applied since
then the denominator can be small. In this case one obtains the law
$v_*=(2^{1/2}\delta)^{1/3}$.

It is also possible to compute the frequency of the small oscillations around the potential minimum, which defines a time scale that could be of the order of oscillations found in the
condensate dynamics due to the competition between nonlocal dispersion and trapping forces
\begin{equation}
\Omega = \sqrt{1 + \frac{3}{v_*^3} +P\frac{\left(3 v_*^2- \delta^2\right)}{(v_*^2+\delta^2)^3}}
\end{equation}
An important feature of this formula is that the frequency grows as $v_* \rightarrow 0$, i. e.
as the range of interactions $\delta$, goes to zero. We will check later the validity of these predictions.

\subsection{3D case: Equilibrium points and minimum width}

Considering again the simplifying assumption that $v_x = v_y = v_z = v$
the dynamical equations in this case are
\begin{equation}
  \frac{d^2v}{d\tau^2} +  v  =
    \frac{1}{v^3} + \frac{Pv}{\left(v^2 + \delta^2\right)^{5/2}}
\end{equation}
and the potential
\begin{equation}
V(v) =  \frac{1}{2} v^2 + \frac{1}{2v^2} + \frac{P}{3\left(v^2+\delta^2\right)^{3/2}}
\end{equation}
As before, collapse can be ruled out since the singularity has been removed. However the equilibrium
point satisfies a more complicated equation
\begin{equation}
\label{equil2}
 v = \frac{1}{v^3} + \frac{Pv}{\left(v^2 + \delta^2\right)^{5/2}}
\end{equation}
Defining again $q=v^2/\delta^2$ the reduced equation is
\begin{equation}
\label{delta2}
\delta^4 = \frac{1}{q^2} + \frac{P}{\delta \left(q + 1\right)^{5/2}}
\end{equation}
and the $\delta$ dependence of $q$ is nontrivial.
Without making any approximations
Eq. (\ref{delta2}) can be written as
\begin{equation}
\label{lapoli}
\delta^2 \left(q+1\right)^5 \left(\delta^4 q^2-1\right)^2 - P^2 q^4 = 0
\end{equation}
whose solutions  for each $(P, \delta)$ pair provide the right
equilibria $v_*$. Now using the $z$-variables as in the previous
subsection one can prove that there exists only one positive root of
Eq. (\ref{lapoli}).

 It is possible to investigate the
orders of the different terms in (\ref{delta2}) and the only simplifying
assumption is that $q \ll 1$, since
we expect now that collapse is stronger than before as is usual in three-dimensional problems.
Using this assumption we find that $\delta/q^2 = -P$ and thus
\begin{equation}
\label{order}
v_* \simeq \frac{{\delta}^{5/4}}{|P|^{1/4}}
\end{equation}
which is smaller than the two dimensional equilibrium width. This is an indication that even though it seems plausible that collapse does not take place the compression of the width due to frustrated collapse process is stronger than that of the two-dimensional case.

\section{Numerical results}
\label{numerics}

 The theoretical analysis of Sects. \ref{infty} and \ref{III} share two common conclusion: (i) There should be a limit on the minimum width of the wavepacket, a fact that could rule out collapse and (ii) there must exist oscillations on  the wave packet. In order to test these results and the other predictions related to the dynamics of the condensate which we have obtained during our approximate variational analysis we have 
integrated numerically Eq. (\ref{GPEredu}). In our numerical simulations we start with Gaussian initial data and then compute the solution using
a symmetrized second order in time Fourier pseudospectral method. Typical grid sizes range from 
128$\times$ 128 to 512$\times$512 points. Simulation times in adimensional units where of the order of 50-100 (integration step $\Delta t=0.01$)
 which correspond to physical values of the order of seconds, which are about the lifetimes of the
condensates and dynamically allow span a time interval in which the essential features are to be captured.

 In our numerical simulations we studied the region of $U$  values contained between the linear case ($U=0$)
and a high value of $U=100$, which is about ten times above the threshold for collapse in the local case \cite{Perez-Garcia98}. In practice we worked with $\delta^2 \in [0.005,0.1]$. 

The main conclusion of our analysis is that up to the precision of the computation we can 
conclude that {\em there is no collapse in all the situations analyzed}. 
 When $\delta$ is further reduced or the nonlinear coefficient $U$ is increased above a certain limit we cannot decide if collapse is possible since then the spatial resolution of the scheme is not enough to know whether collapse is real or a mere numerical artifact. This is a problem of the numerical simulation which is quite difficult to avoid since as will be discussed in detail later one needs at the same time
a large integration region to capture the global wavefunction shape and a fine grid to resolve
the small amplitude peaks thus making the problem computationally hard. On the other hand, it is not easy to consider the radially symmetric geometry as it is done in local collapse problems \cite{PP2,Ot2,Pi2}.

\begin{figure}
\epsfig{file=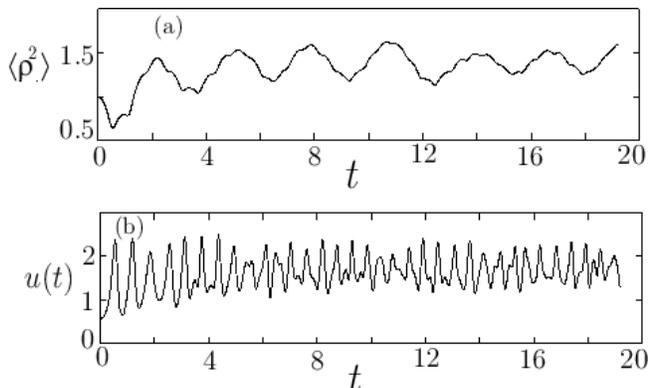,width=8.5cm}
\caption{Width (a) and amplitude (b) oscillations in a supercritical condensate with $U=-20$, $\delta^2 = 0.05$. The frustrated collapse events appear as peaks in the condensate amplitude (maximum height).}
\label{collapso}
\end{figure}

 Results of a typical simulation are shown in detail in Fig. \ref{collapso}. Two things are clear: firstly that the wave packet width oscillates with a minimum width of order ${\cal O}(1)$, secondly the maximum amplitude of the condensate performs oscillations with a dominant frequency.

\begin{figure}
\epsfig{file=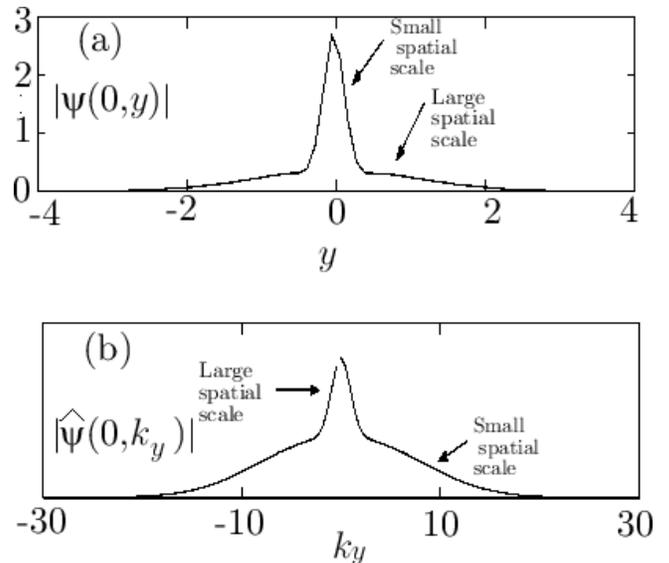,width=8.5cm}
\caption{(a) Trasversal section of the condensate amplitude $|\psi(0,y)|$ during a frustrated collapse event for $t \simeq 0.4$. The two contributions to the solution are very clear. The spatial spectrum (b) has also signatures of the existence of two scales.}
\label{dospartes}
\end{figure}

 One could expect that, according to Eq. (\ref{cero}), the minimum width of the wavepacket were
of order $\delta$. However that would be the case if all the wave function were involved in the frustrated collapse events, which is not the case. As it happens in collapse in the local Nonlinear Schr\"odinger equation only part of the wave function takes part on the squeezing dynamics, the remaining part being the responsible of the finite size width. In our case this is very clear in Fig. \ref{dospartes} where the two contributions to the solution are clearly seen, one being of order ${\cal O}(1)$ and the other corresponding to a smaller scale $\ell_c$. Also it is interesting how the low amplitude extended (``inert part")
oscillates according to trap frequency while the peak dynamics is ruled out by the nonlocal dynamics, a phenomenon which is clear in the width oscillations of Fig. \ref{collapso}(a), were the two frequencies are present in the dynamics. In fact, the local oscillations were predicted in Sec. \ref{infty} and appear also as oscillations in the potential well in the variational formalism. 
 They correspond to ``frustrated collapse" events since if the nonlocality were not present the concentration dynamics would not stop at scale $\ell_c$ but would continue up to infinity.

\begin{figure}
\epsfig{file=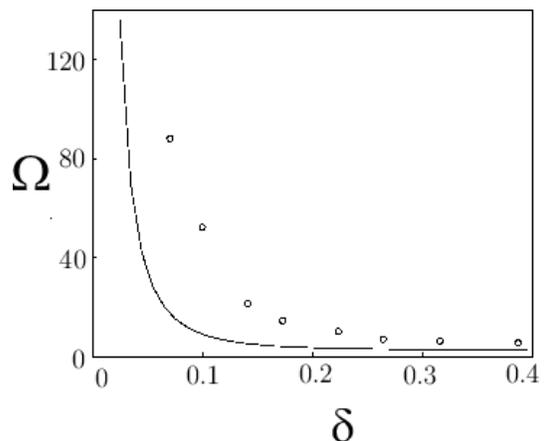,width=7.0cm}
\caption{Oscillation frequency of the condensate (circles) for $U=-20$ and different 
$\delta$ values against the variational estimate (solid line).}
\label{ondelta}
\end{figure}

It is remarkable that even when the variational method does not take into account the 
existence of two scales at least some of its predictions  
match reasonably well with the  simulated dynamics of  
Eq. (\ref{GPEredu}). For example Fig. \ref{ondelta} shows the frequency of the amplitude oscillations of the condensate  as a function of $\delta$ (circles). The variational estimates (solid line) 
are in qualitative agreement. This result cannot be improved in the framework of simple variational analysis since it corresponds to the Gaussian ansatz and is derived for the oscillations near the bottom of the potential well.

\begin{figure}
\epsfig{file=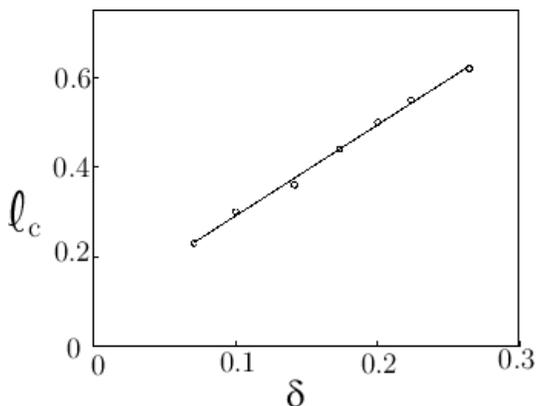,width=7.0cm}
\caption{Minimum width of the ``collapsing" part $\ell_c$ (width of the small scale peak) as a function of $\delta$ for $U = -20$. The circles represent the values obtained by the numerical simulations of Eq. (\ref{GPEredu}) and the solid line is a least squares interpolant, which proves the linear dependence on $\delta$.}
\label{comparison}
\end{figure}

Other result of the variational analysis which is quite relevant is the size of the small scale
$\ell_c$  generated during the partial compression of the wave packet, which should be of order $\delta$ for a two dimensional condensate. To check it we 
have estimated the width of the collapsing peak using the small scale 
of the spatial representation of $\psi$ plot and the long scale of spatial spectra for different values of $\delta$. Our results are summarized in Fig. \ref{comparison}, were 
it can be seen how the width of the peak depends linearly on $\delta$, which again supports the variational analysis.

\section{Conclusions}
\label{IV}

 In this paper we have presented analytical and numerical evidences 
that the effect of nonlocal interactions is to prevent collapse when this is the dominant 
physical interaction process in negative scattering 
length Bose-Einstein condensates. The analytical tools 
combine the exact analysis of the highly nonlocal limit,
 moment analysis and collective coordinate analysis. Even
 though there is no a rigorous proof, the predictions of all 
methods point out that collapse is excluded when nonlocal (and  
nonsingular) interactions are taken into account. 
 
Most of the analytical predictions have been based upon the assumption of cylindrical or spherical symmetry of the wavefunction although it is easy to generalize the results to other symmetries. From the numerical solutions one sees that radially symmetric solutions are stable and no asymmetric instabilities grow when one starts with symmetric initial data. In fact, in the context of local GP equation collapse phenomenon seems to be an essentially radially symmetric process \cite{Weinstein,Qi2}.

The analytical predictions have been tested with a numerical scheme. Up to the range in which the numerical simulations can be trusted we have not observed collapse in the system which supports our theoretical analysis, even when the last concentrates on the case
when all the solution collapses (both the strongly nonlocal limit and the variational method).

 For BEC the main implication is that if one were able to generate large negative scattering length condensates as discussed in the introduction, i. e. by overlapping of two subcritical ones or by switching the scattering length from positive to negative, one could obtain an oscillating but non collapsing condensate provided there were no additional physical effects to be taken into account. In fact since it seems that the Hamiltonian is bounded below one could 
construct a ground state for this system so that stationary states can probably be obtained.

From a fundamental point of view our analysis is the first systematic study of the behavior of the effect of nonlocal interacting Bose-Einstein condensates taking into account different aspects: variational analysis, moment approximations and numerical techniques. We hope this study will be valuable in the search for stable attractively interacting condensates.

\acknowledgements

Authors are indebt to G. Alfimov for many fruitful discussions and 
a critical reading of the manuscript.
VMPG has been partially supported by Spanish DGCYT
(grant PB96-0534).
The work of VVK has been supported by FEDER and by the Program PRAXIS
XXI, grant No.  PRAXIS/2/2.1/FIS/176/94.


\begin{references}
\bibitem{com}{It is to be mentioned that would the collapse occcur 
the mentioned assumption would mean rather high densities of the 
condensate. Taking the range of interactions to be of order of 
$10^3\times a_B$ where $a_B$ is the Bohr radis \cite{Parola} and 
assuming that about $10^3$ atoms are involved in the region of 
collapse, the density of particles is estimated to be $10^{18}$ 
cm$^{-3}$. Meantime, the purpose of the mentioned assumption is to 
show that it results in spreading out of the wave packet and thus is 
roughly  speaking not reachable even formally, i. e. with the 
framework of the nonlocal GP equation.}
\bibitem{phys}{This approximation is equivalent to saying that at the scale at wich nonlocal interactions become relevant the trap variations are quite small}
\bibitem{DIsc}{This is a characteristic of collapses in Nonlinear Schr\"odinger systems which makes the analysis difficult. Even for the local interaction case there has been controversy on how the stationary and collapsing parts interact during the collapse process. See e. g. the different proposals in \onlinecite{Ot2,Pi2}}
\end{references}
 \end{document}